# Detection opportunity for aromatic signature in Titan's aerosols in the 4.1 to 5.3 μm range

Christophe Mathé[1], Thomas Gautier[2], Melissa G. Trainer[4], Nathalie Carrasco[2,3]

[1]LESIA, Observatoire de Paris, PSL Research University, CNRS, Sorbonne Universités, UPMC Univ. Paris 06, Univ. Paris Diderot, Sorbonne Paris Cité, 5 place Jules Janssen 92195 MEUDON

[2]LATMOS, CNRS, UVSQ, UPMC, UPSaclay, 11 Bvd d'Alembert 78280 Guyancourt, France

[3]Institut Universitaire de France,

[4]NASA Goddard Space Flight Center, 8800 Greenbelt Rd, Greenbelt MD, USA

**Running head:** New observation window for Titan aerosols in the IR

**Corresponding authors:**

christophe.mathe@obspm.fr ;

thomas.gautier@latmos.ipsl.fr

# ABSTRACT

The Cassini/Huygens mission provided new insights on the chemistry of the upper atmosphere of Titan. The presence of large molecules and ions (>100's of amu) detected by Cassini at high altitude was not expected, and questions the original assumptions regarding the aerosol formation pathways. From recent laboratory studies, it has been shown that the inclusion of trace amounts of aromatic species drastically impacts the chemistry of aerosol formation and induces observable changes in the properties of the aerosols. In the present work we focus on the effect of one of the simplest nitrogenous aromatics, pyridine ($C_5H_5N$), on the infrared signature of Titan's aerosol analogs. We introduce initial gas mixtures of: (i) $N_2$-$C_5H_5N$ (100%/250ppm), (ii) $N_2$-$CH_4$-$C_5H_5N$ (99%/1%/250ppm), (iii) $N_2$-$CH_4$ (99%/1%) in a cold plasma discharge. The material produced, herein called tholins, is then analyzed by mid-infrared spectroscopy. When adding pyridine in the discharge, the tholins produced present an aromatic signature in the 4.1-5.3 µm (1850-2450 $cm^{-1}$) spectral region, attributed to overtones of aromatic C-H out-of-plane bending vibrations. We also observe a spectral shift of the nitrile and iso-nitrile absorption band with the inclusion of pyridine in the gas mixture. These results could help to investigate the data obtained at Titan by the Cassini/VIMS instrument in the 1-5 µm infrared window.

# 1 INTRODUCTION

The largest moon of Saturn, Titan, has fascinated numerous scientists because it is the only moon in our solar system that possesses an atmosphere. Titan's atmosphere is composed of nitrogen (~98%), methane (~1.5%) and traces species of hydrocarbons (Waite et al., 2005). In the ionosphere (~1000 km), gas is dissociated and ionized by solar photon irradiation and energetic particles from Saturn's magnetosphere. This chemistry leads to the formation of complex photochemical aerosols in the atmosphere. Since the discovery of Titan's thick haze layer by the Voyager probe (Smith et al, 1981), the question of the aerosol composition has been widely investigated (Gautier et al., 2012; Imanaka et al., 2004; Khare et al., 1984). The results from the Cassini/Huygens mission have provided constraints on the chemical composition of the aerosol, though not a comprehensive analysis (Hörst, 2017). Results from the Huygens/ACP experiment demonstrated that the atmospheric aerosols incorporate a large fraction of nitrogen (Israël et al., 2005). Cassini/CIRS (Composite Infrared Spectrometer) detected C-H bending mode in the aerosol signature in the far-infrared (Vinatier et al., 2012).

However, aromaticity in Titan's aerosol remains uncertain. Dinelli et al. (2013) reported a detection of an unidentified emission around 3.28 µm (3049 $cm^{-1}$) in Titan's atmosphere, observed by the VIMS (Visual and Infrared Mapping Spectrometer) instrument aboard Cassini. Based on the in situ measurements of small molecules by INMS (Ion and Neutral Mass Spectrometer), the feature could be attributed to aromatic molecules, like benzene (Waite et al., 2007). Yet, the small concentration of benzene (few ppm) did not explain the large magnitude of the emission peak observed by VIMS. Dinelli et al. (2013) suggested that the unidentified emission could be due to a solid signature of polycyclic aromatic hydrocarbons (PAHs) or heterocyclic aromatic compounds in the Titan's upper atmosphere (Coates et al., 2007; Waite et al., 2007). López-Puertas et al. (2013) further attempted to compare this unidentified emission peak to PAH emission features. PAHs have their vibrational signatures in the spectral range of the unidentified emission. Moreover, PAHs excited by UV-solar radiation redistribute internally the energy in their vibrational modes causing near and mid-IR emissions. To analyze this unidentified emission, VIMS observations were compared to an excitation model of a set of PAHs and nitrogenous PANHs. The observed spectral feature from VIMS was found to be consistent with the combination of several PAH spectra. Lopez-Puertas et al. (2013) established a list of the nineteen most abundant PAHs fitting well with the VIMS observations and composed of 10-11 aromatics rings on average. This attribution is not unique or univocal, but provides strong evidence that there may be an aromatic component in Titan's aerosols.

In the present work, we investigate the infrared spectra of aerosol analogs with a high aromatic content with the aim of identifying new features that would be indicators the aromatic component of Titan's aerosols. To produce such highly aromatic aerosols, we begin with trace amounts of pyridine ($C_5H_5N$). Indeed, previous laboratory investigations by Sebree et al. (2014) showed that aerosols produced with trace amounts of 1- and 2-ring aromatics provided the best match to broad aerosol emission features below 200 $cm^{-1}$ actually, observed at Titan in CIRS far-infrared spectra (Anderson & Samuelson, 2011). Aerosol analogs with a nitrogen component, including those formed with pyridine, provided the strongest feature in this region of the spectrum. Here, pyridine is chosen in order to favor a high nitrogen content, consistently with the Huygens/ACP detection (Israel et al. 2005), and because this compound has been inferred from INMS data to be present in the upper atmosphere (Vuitton et al., 2007).

A typical signature for aromaticity would be the fingerprints due to the $\nu_{(0-1)}$ transition of the out-of-plane C-H bending modes, which are usually located in the 600-900 $cm^{-1}$ wavenumber range (Tommasini et al., 2016). However, this region is generally crowded for hydrocarbons (Coustenis et al., 2010), making it challenging to decipher in complicated spectra. Fortunately such fingerprints have a counterbalance due to aromatic overtones linked to the $\nu_{(0\text{-}3)}$ transition in the 1800-2100 $cm^{-1}$ wavenumber range, which happens to be a region of the spectrum free from any other vibrational absorptions (Herzberg, 1950). We thus choose to investigate this spectral band, which partially coincides with a Cassini/VIMS observable range (1 to 5 µm) (Brown et al., 2005), to characterize the spectra of our laboratory aerosols (Brown et al., 2005).

## 2 EXPERIMENTAL DETAILS

Aerosols analogs were produced using the PAMPRE experiment, based on a plasma discharge in a gas mixture of nitrogen with methane and/or pyridine. Then, the tholins produced are extracted from the reactor to be analyzed by infrared spectroscopy with an attenuated total reflection method.

The PAMPRE setup is based on a capacitively coupled radio-frequency (13.6 MHz) system as energy source to induce a plasma in gas mixtures. The experimental details can be found in Szopa et al., (2006).

The gas mixture is injected into the top of the vacuum reactor at a 55 sccm (standard cubic centimeter per second) flow rate. The PAMPRE reactor is operated at room temperature, at a pressure of 0.9 mbar and an effective power of 30 W.

When the reactor is on, the plasma induces chemistry that leads to the formation of organic aerosol, considered analogs of Titan's haze particulates. Due to charge effects, aerosols are maintained in levitation in the plasma allowing for growth in the free atmosphere without induced wall effects. The aerosol formation process is self-limiting. Once the solid particles reach a critical diameter they become too heavy, and the electric field yields and the gravitation forces take over. The particles fall through the grid and deposit outside the reactor for collection on a glass vessel. Theses solid products are called tholins. An estimate of the production rate, defined as the ratio between the mass of tholins produced and the production duration, is obtained by weighing the total collected sample with a 0.1 mg precision scale. This estimate only correspond to the lower limit of the production rate as a few tholins remains on the glass vessel due to electrostatic charging and cannot be extracted. The detailed protocol and a discussion on the limitations of this procedure together with a study on the influence of methane percentage on tholins production rate in the PAMPRE reactor can be find in Sciamma-O'Brien et al. (2010)□.

For this work, three different gas mixtures were used to produce aerosols as presented in table 1: (1) a reference case with only $N_2$ and $CH_4$ in the initial gas mixture, (2) a second case to evaluate the effect of traces of pyridine in addition to $CH_4$ to compare to the reference mix, and (3) a third case to evaluate the capacity of pyridine itself to generate aerosols with a highly aromatic signature.

To test the effect of pyridine in the $N_2$-$CH_4$ chemical system, we have chosen to set the methane concentration at 1% in our experiments, in agreement with a lower limit for Titan's atmospheric composition (Niemann et al., 2010)□.

Table 1

| *Initial Gas mixture* | *$N_2$ (%)* | *$CH_4$ (%)* | *$C_5H_5N$ (ppm)* |
|---|---|---|---|
| *$N_2$-$CH_4$* | 99 | 1 | 0 |
| *$N_2$-$CH_4$- $C_5H_5N$* | ~99 | 1 | 250 |
| *$N_2$- $C_5H_5N$* | ~100 | 0 | 250 |

*Table 1: Initial composition of the gas mixtures.*

Experiments were made using reactants from Air Liquide: $N_2$ (>99,999% purity), a certified $CH_4/N_2$ mixture (10% $CH_4$, > 99,999 % purity) and a certified pyridine/$N_2$ mixture (550 ppm of pyridine, > 95 % purity).

The ATR (Attenuated Total Reflection) method enable a direct measurement of the infrared spectra of solid product without further preparation. For this analysis a few micrograms of bulk tholins are deposited onto the ATR plate of a Nicolet 6700 Fourier Transform InfraRed - Attenuated Total Reflection from Thermo Fisher Scientific. Spectra were recorded on the 1800-2500 $cm^{-1}$ wavenumber range with a 2 $cm^{-1}$ spectral resolution and a Michelson velocity set at 0.624 $cm^{-1}/s$. All spectra were normalized to the most intense peak around 2169 $cm^{-1}$ and corrected for atmospheric background.

# 3 RESULTS AND DISCUSSION

## 3.1 Aerosol Production Efficiency

Rough estimate of the production rates were measured for the two experiments performed with 250 ppm of pyridine, with and without methane. For a similar plasma duration of 6 hrs, we found a mass production of 70 and 8.2 mg with and without methane respectively. The production rate in the presence of 1% methane is an order of magnitude larger (11.7 mg/h and 1.4 mg/h respectively), but needs to be scaled to consider the relative amount of carbon introduced in the two experiments. Each molecule of pyridine contains 5 carbon atoms, so the 250 ppm of pyridine provides 1250 ppm carbon atoms to the reactive mixture. There is therefore one order of magnitude less carbon provided by 250 ppm pyridine than by 1% methane amount in the discharge. Thus, the methane and the aromatic pyridine have roughly the same efficiency to produce solid particles in the plasma experiment. This suggests that on Titan a few ppm of aromatic molecules could contribute significantly to the aerosol formation, especially at altitudes where methane photochemistry is no longer active. This is consistent with previously reported studies (Trainer et al. 2013, Sebree et al 2014).

## 3.2 Aerosol Spectral Signature

The spectra of the three samples in the 1850 to 2450 $cm^{-1}$ range are shown in figure 1.

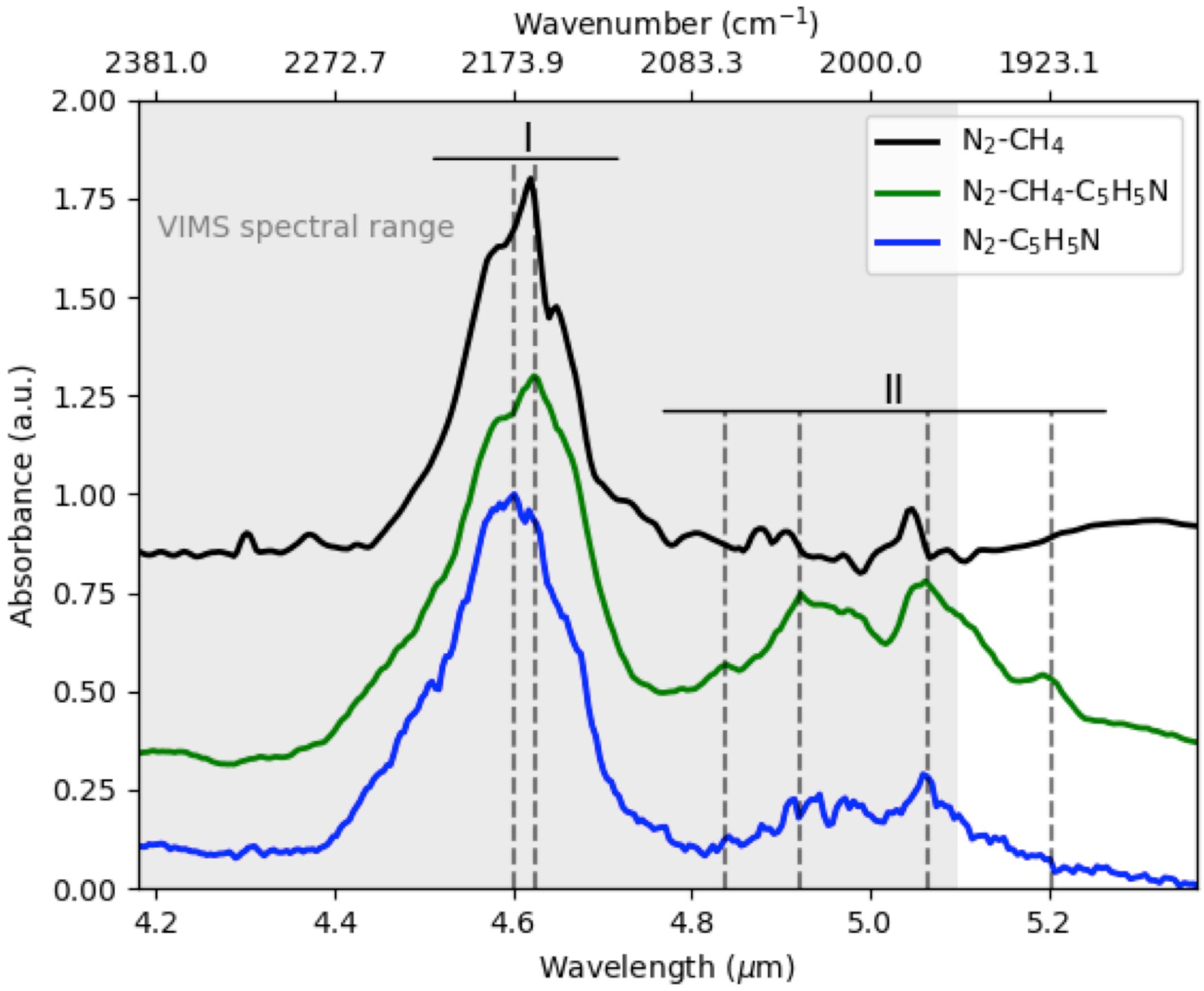

*Figure 1: Infrared spectra of tholins in the 4.1-5.3 µm spectral range (1850-2450 cm$^{-1}$). The curves are labeled with the initial gas mixture used to produce tholins (see table 1): $N_2$-$CH_4$ (black), $N_2$-$C_5H_5N$ (blue) and $N_2$-$CH_4$-$C_5H_5N$ (green). Regions I and II correspond to nitriles vibration and aromatic overtones respectively. The gray corresponds to the VIMS spectral range.*

Two features stand out in this range of the spectra: one region between 2100 and 2250 cm$^{-1}$ (zone I), and the second between 1900 and 2100 cm$^{-1}$ (zone II). For each region, the band center position, intensity and potential attribution are detailed in table 2.

*Table 2*

| Region | $N_2$-$CH_4$ | | $N_2$-$C_5H_5N$ | | $N_2$-$CH_4$-$C_5H_5N$ | | Potential attributions |
|---|---|---|---|---|---|---|---|
| | *Center μm ($cm^{-1}$)* | *Intensity* | *Center μm ($cm^{-1}$)* | *Intensity* | *Center μm ($cm^{-1}$)* | *Intensity* | |
| I | 4.65 (2151) | m | 4.68 (2137) | m | 4.66 (2145) | w | *-N=C=N- (l)*<br>*-CN (m)* |
| | 4.62 (2164) | vs | 4.60 (2173) | vs | 4.62 (2163) | vs | *Conjugated -CN, such as -C=C(-NH2) (l)*<br>*C-NC stretching (l)*<br>*-NC (m)*<br>*$C_{ar}$-CN ?* |
| | - | - | 4.52 (2210) | w | 4.54 (2204) | w | *$C_{ar}$-NC* |
| | 4.48 (2230) | w | 4.48 (2233) | m | 4.47 (2235) | w | *R-NC stretching (l)*<br>*-N=C=N- (m)* |
| II | - | - | - | - | 5.20 (1922) | m | *Overtone C-H (solo)* |
| | 5.05 (1980) | vw | 5.06 (1977) | w | 5.06 (1975) | s | *Overtone C-H (duo)* |
| | - | - | 4.94 | w | 4.92 | s | *Overtone C-H (trio)* |
| | | | (2025) | | (2032) | | |
| | - | - | 4.89 (2045) | vw | 4.84 (2067) | w | *Overtone C-H (quatro)* |

*Table 2: List of the absorption bands detected in the 4.1-5.3 µm spectral range.*

**Note.** Potential attribution of peaks in regions I are taken from (1) Imanaka et al (2004) and (2) Gautier et al (2012). Band intensity legend: *vs* = very strong; s = strong, m = medium, *w* = weak, *vw* = very weak.

Zone I corresponds to the well-known absorption bands due to the nitrile (R-CN) and iso-nitrile (R-NC) functions (Gautier et al., 2012; Imanaka et al., 2004)□. When comparing the three spectra in figure 1, the first result is the similarity in position of this feature. It is dominated by an absorption band peaking at 2173 $cm^{-1}$ corresponding to the stretching mode of nitrile or iso-nitrile terminal functions branched on aliphatic structures.

However, a second contribution appears in this broad band, peaking at around 2240 $cm^{-1}$ (see peaks attribution in table 2). The weight of this contribution is strongest in the $N_2$-$C_5H_5N$ spectrum, becoming a clear shoulder on the band at 2173 $cm^{-1}$ (blue curve in figure 1). The shift to a higher wavenumber is consistent with the signature of nitrile functionalities branched on aromatic rings (Sciamma-O'Brien et al., 2017)□.

The absorption features in zone I are consistent with the formation of nitrile and iso-nitrile functions, primarily branched from aliphatic structures, for the three tholins samples - even in the experiment with only pyridine as the carbon source. However, some of the nitrile and iso-nitrile functional groups may also be branched from aromatic rings, as suggested with the contribution of a signature at around 2240 $cm^{-1}$. This signature is especially prominent in the samples produced with pyridine.

Zone II shows a new band appearing between 1900-2100 $cm^{-1}$ in the spectra of samples produced with pyridine. Four mains peaks can be distinguished within the $N_2$-$CH_4$-$C_5H_5N$ (green curve) at positions: 1922, 1973, 2026, 2074 $cm^{-1}$. These bands have also been observed in the recent work from (Sciamma-O'Brien et al., 2017)□ in tholins produced with mixtures of nitrogen, methane and benzene.

This spectral region does not match with any fundamental vibration of organic chemical functional groups, but we can confidently attribute them to aromatic derivatives overtones ($\nu_{0-3}$) since:

1/ A recent simulation study on the fundamental fingerprints ($\nu_{0-1}$) of 51 pure PAHs has shown the spectral region between 730-910 $cm^{-1}$ corresponds to out-of-plane CH bending at the edge of the aromatic cycle (Tommasini et al., 2016)□,

2/ Overtone signatures are due to the *n* harmonic of the fundamental vibrational state ($\nu_{0-1}$) of a molecule, corresponding to the $\nu_{0-n}$ transition (Kendall 1966, Young et al., 1951)□,

3/ The frequencies of overtones do not correspond exactly to *n*-times the frequency of the fundamental vibration, but rather are expected to show some shift toward lower wavenumber due to the anharmonicity of molecules (Herzberg 1950),

And,

4/ Gautier et al. (2017)□ also reported the detection of polymeric patterns in the mass spectra of tholins produced with gas mixture of nitrogen, methane and one or two rings aromatics. These were interpreted as a possible signature of aromatic functions within the structure of tholins produced in such gas mixture.

The detection of theses C-H out-of-plane bending at the edge of the aromatic cycle overtones, in a region free of fundamental vibration of any other organic chemical functional groups, confirms the inclusion of aromatic cycles in the macromolecular structure of our tholins and provides a first observational opportunity for the determination of aromatic content of Titan's aerosols.

Thus we suggest that the observation of these bands, which could be partially detectable in the Cassini/VIMS-IR observational range, would provide the strongest indication yet regarding the aromaticity of Titan's aerosols.

# 4 Conclusion

In the present work we identified a strongly indicative signature of aromaticity of Titan's aerosol analogs in the 4.3-5.1 µm spectral range. This signature is attributed to overtones of the aromatic structural vibrations of the aerosols. The latter frequencies correspond to out-of-plane bending vibrations in the 11.7-15.4 µm spectral range. Two instruments on Cassini have been able to observe the Titan atmosphere at these wavelengths. The fingerprint frequencies would be in the FP3 window of the CIRS instrument, and were tentatively reported by (Vinatier et al., 2012) with a signature at 13.25 µm. The corresponding overtones would be within the VIMS-IR spectral range.

Regarding a search for an aromatic signature in Titan's aerosols, our results indicate that targeting the overtones in the existing VIMS-IR data and the fingerprint frequencies in the CIRS-FP3 window would provide a strategy for complementary detections.

Our work therefore provides a new detection opportunity for aromatic signature in Titan's aerosols using the 4.1 - 5.3 µm range of the VIMS-IR instrument. Such a detection could provide the most convincing evidence of the possible aromaticity of Titan's aerosols. The aromatic or aliphatic nature of Titan's aerosols is an important issue, as these functionalities affect optical properties and chemical activity. A significant aromatic component in the aerosol would induce a different fate of the aerosols both in the atmosphere and on Titan's surface, for example regarding their reactivity at low altitude or on the surface (Gudipati et al., 2013) and their respective solubility in Titan's lakes (Cable et al., 2012; Chevrier et al., 2015; Malaska et al., 2014).

## Acknowledgements

C.M. wishes to thank the Paris Observatory for it financial support for this work. N.C. thanks the European Research Council for its financial support under the ERC PrimChem project ( grant agreement No. 636829). M.G.T thanks the NASA Solar System Workings Program for financial support. The authors thank Fabrice Duvernay for the fruitful discussion on the anharmonicity of overtones signatures.

## References

Anderson, C. M., & Samuelson, R. E. (2011). Titan's aerosol and stratospheric ice opacities between 18 and 500 μm: Vertical and spectral characteristics from Cassini CIRS. *Icarus*, *212*(2), 762–778. https://doi.org/10.1016/j.icarus.2011.01.024

Brown, R. H., Baines, K. H., Bellucci, G., Bibring, J. P., Buratti, B. J., Capaccioni, F., … Sotin, C. (2005). The Cassini visual and infrared mapping spectrometer (VIMS) investigation. *Space Science Reviews*, *115*(1–4), 111–168. https://doi.org/10.1007/s11214-004-1453-x

Cable, M. L., Hodyss, R., Beauchamp, P. M., Smith, M. A., & Willis, P. A. (2012). Titan Tholins: Simulating Titan Organic Chemistry in the Cassini-Huygens Era, 1882–1909. https://doi.org/10.1021/cr200221x

Chevrier, V. F., Luspay-Kuti, A., & Singh, S. (2015). Experimental study of nitrogen dissolution in methane-ethane mixtures under Titan surface conditions. *LPSC*, 12–13.

Coates, A. J., Crary, F. J., Lewis, G. R., Young, D. T., Waite, J. H., & Sittler, E. C. (2007). Discovery of heavy negative ions in Titan's ionosphere, *34*(June), 6–11. https://doi.org/10.1029/2007GL030978

Coustenis, A., Jennings, D. E., Nixon, C. A., Achterberg, R. K., Lavvas, P., Vinatier, S., … Romani, P. N. (2010). Titan trace gaseous composition from CIRS at the end of the Cassini-Huygens prime mission. *Icarus*, *207*(1), 461–476. https://doi.org/10.1016/j.icarus.2009.11.027

Dinelli, B. M., Lõpez-Puertas, M., Adriani, A., Moriconi, M. L., Funke, B., García-Comas, M., & D'Aversa, E. (2013). An unidentified emission in Titan's upper atmosphere. *Geophysical Research Letters*, *40*(8), 1489–1493. https://doi.org/10.1002/grl.50332

Gautier, T., Carrasco, N., Mahjoub, A., Vinatier, S., Giuliani, A., Szopa, C., … Cernogora, G. (2012). Mid- and far-infrared absorption spectroscopy of Titan's aerosols analogues. *Icarus*, *221*(1), 320–327. https://doi.org/10.1016/j.icarus.2012.07.025

Gautier, T., Sebree, J. A., Li, X., Pinnick, V. T., Grubisic, A., Loeffler, M. J., … Brinckerhoff, W. B. (2017). Influence of trace aromatics on the chemical growth mechanisms of Titan aerosol analogues. *Planetary and Space Science*, *140*(November 2016), 27–34. https://doi.org/10.1016/j.pss.2017.03.012

Gudipati, M. S., Jacovi, R., Couturier-tamburelli, I., Lignell, A., & Allen, M. (2013). Photochemical activity of Titan's low altitude condensed haze. *Nature Communications*, *4*(April), 1648. https://doi.org/10.1038/ncomms2649

Herzberg, 'Molecular spectra and molecular structure', 1950

Hörst, S. M. (2017). Titan's atmosphere and climate. *Journal of Geophysical Research: Planets*, *122*(3), 432–482. https://doi.org/10.1002/2016JE005240

Imanaka, H., Khare, B. N., Elsila, J. E., Bakes, E. L. O., McKay, C. P., Cruikshank, D. P., ... Zare, R. N. (2004). Laboratory experiments of Titan tholin formed in cold plasma at various pressures: Implications for nitrogen-containing polycyclic aromatic compounds in Titan haze. *Icarus*, *168*(2), 344–366. https://doi.org/10.1016/j.icarus.2003.12.014

Israël, G., Szopa, C., Raulin, F., Cabane, M., Niemann, H. B., Atreya, S. K., ... Vidal-Madjar, C. (2005). Complex organic matter in Titan's atmospheric aerosols from in situ pyrolysis and analysis. *Nature*, *438*(7069), 796–799. https://doi.org/10.1038/nature04349

Kendall, 'Applied Infrared spectroscopy', 1966

Khare, B. N., Sagan, C., Arakawa, E. T., Suits, F., Callcott, T. A., & Williams, M. W. (1984). Optical constants of organic tholins produced in a simulated Titanian atmosphere: From soft x-ray to microwave frequencies. *Icarus*, *60*(1), 127–137. https://doi.org/10.1016/0019-1035(84)90142-8

López-Puertas, M., Dinelli, B. M., Adriani, A., Funke, B., García-Comas, M., Moriconi, M. L., ... Allamandola, L. J. (2013). Large abundances of polycyclic aromatic hydrocarbons in Titan's upper atmosphere. *Astrophysical Journal*, *770*(2). https://doi.org/10.1088/0004-637X/770/2/132

Malaska, M. J., & Hodyss, R. (2014). Dissolution of benzene, naphthalene, and biphenyl in a simulated Titan lake. *ICARUS*, *242*, 74–81. https://doi.org/10.1016/j.icarus.2014.07.022

Niemann, H. B., Atreya, S. K., Demick, J. E., Gautier, D., Haberman, J. A., Harpold, D. N., ... Raulin, F. (2010). Composition of Titan's lower atmosphere and simple surface volatiles as measured by the Cassini-Huygens probe gas chromatograph mass spectrometer experiment. *Journal of Geophysical Research E: Planets*, *115*(12), 1–22. https://doi.org/10.1029/2010JE003659

Sciamma-O'Brien, E., Carrasco, N., Szopa, C., Buch, A., & Cernogora, G. (2010). Titan's atmosphere: An optimal gas mixture for aerosol production? *Icarus*, *209*(2), 704–714. https://doi.org/10.1016/j.icarus.2010.04.009

Sciamma-O'Brien, E., Upton, K. T., & Salama, F. (2017). The Titan Haze Simulation (THS) experiment on COSmIC. Part II. Ex-situ analysis of aerosols produced at low temperature. *Icarus*, *289*, 214–226. https://doi.org/10.1016/j.icarus.2017.02.004

Sebree, J. A., Trainer, M. G., Loeffler, M. J., & Anderson, C. M. (2014). Titan aerosol analog absorption features produced from aromatics in the far infrared. *Icarus*, *236*, 146–152. https://doi.org/10.1016/j.icarus.2014.03.039

Smith, Bradford A., et al. "Encounter with Saturn: Voyager 1 Imaging Science Results." *Science*, vol. 212, no. 4491, 1981, pp. 163–191.

Szopa, C., Cernogora, G., Boufendi, L., Correia, J. J., & Coll, P. (2006). PAMPRE: A dusty plasma experiment for Titan's tholins production and study. *Planetary and Space Science*, *54*, 394–404. https://doi.org/10.1016/j.pss.2005.12.012

Tommasini, M., Lucotti, A., Alfè, M., Ciajolo, A., & Zerbi, G. (2016). Fingerprints of polycyclic aromatic hydrocarbons (PAHs) in infrared absorption spectroscopy. *Spectrochimica Acta - Part A: Molecular and Biomolecular Spectroscopy*, *152*, 134–148. https://doi.org/10.1016/j.saa.2015.07.070

Vinatier, S., Rannou, P., Anderson, C. M., Bézard, B., De Kok, R., & Samuelson, R. E. (2012). Optical constants of Titan's stratospheric aerosols in the 70-1500cm-1spectral range constrained by Cassini/CIRS observations. *Icarus*, *219*(1), 5–12. https://doi.org/10.1016/j.icarus.2012.02.009

Vuitton, V., Yelle, R. V, & Mcewan, M. J. (2007). Ion chemistry and N-containing molecules in Titan's upper atmosphere, *191*, 722–742. https://doi.org/10.1016/j.icarus.2007.06.023

Waite, J. H., Niemann, H. B., Yelle, R. V., Kasprzak, W. T., Cravens, T. E., & Luhmann, J. G. (2005). Ion Neutral Mass Spectrometer Results from the First Flyby of Titan. *Science*, *308*, 982–986.

Waite, J. H., Young, D. T., Cravens, T. E., Coates, A. J., Crary, F. J., Magee, B., & Westlake, J. (2007). Planetary science: The process of tholin formation in Titan's upper atmosphere. *Science*, *316*(5826), 870–875. https://doi.org/10.1126/science.1139727

Young, C. W., Du Vall, R. B., & Wright, N. (1951). Characterization of benzene ring substitution by Infrared SpectraYoung, C. W., Du Vall, R. B., & Wright, N. (1951). Characterization of benzene ring substitution by Infrared Spectra. Analytical Chemistry, 2(5), 1–6. https://doi.org/10.1021/ac60053a007. *Analytical Chemistry*, *2*(5), 1–6. https://doi.org/10.1021/ac60053a007